\documentclass[cameraready]{Interspeech}

\title{Hierarchical Permutation Consistency Learning for Self-Conditioned End-to-End Speaker Diarization}

\author[affiliation={1}, orcid=0009-0009-7742-2846]{Bongsu}{Jung}
\renewcommand{\authorsep}{ and } 
\author[affiliation={1,2}, orcid=0000-0002-3854-2783]{Wooil}{Kim}

\address{
    $^1$ Department of Computer Science and Engineering, Incheon National University\\ $^2$ Human-centered Artificial Intelligence Research Center, Incheon National University, Incheon 22012, Republic of Korea
}

\email{\{qhd0330, wikim\}@inu.ac.kr}

\keywords{speaker diarization, end-to-end neural diarization (EEND), self-conditioning, permutation invariant training (PIT), hierarchical permutation consistency}

\usepackage{comment}
\usepackage{multirow} 
\usepackage{booktabs}

\begin{document}

\maketitle

\begin{abstract}
    Self-conditioned (SC) end-to-end neural diarization (EEND) iteratively refines speaker predictions across layers. However, conventional layer-wise permutation invariant training (PIT) independently determines the optimal speaker permutation at each layer. This layer-specific permutation selection causes hierarchical permutation inconsistency (HPI), propagating permutation noise into the SC path and disrupting training stability. To resolve this, we propose two frameworks: Group-Wise PIT (GW-PIT) and Continuous Influence PIT (CI-PIT). While GW-PIT enforces a global permutation constraint, CI-PIT applies soft regularization by aggregating layer-wise cost matrices using Gaussian weighting. This promotes permutation consistency across the hierarchy while preserving local representation flexibility. Experimental results on CALLHOME demonstrate that the proposed CI-PIT suppresses permutation noise, stabilizes training dynamics, and achieves DER improvements for 2- and 3-speaker scenarios.
\end{abstract}

\section{Introduction}
\label{sec:intro}

Speaker diarization, the task of identifying ``who spoke when,'' is a core front-end for various speech applications~\cite{park2022review}. To address the error propagation and structural limitations of traditional clustering pipelines in overlapping speech~\cite{park2022review}, End-to-End Neural Diarization (EEND) reformulated the task as a frame-wise multi-label classification problem~\cite{fujita2019end_1}. EEND architectures have since evolved from self-attention~\cite{fujita2019end} and encoder-decoder attractors (EDA)~\cite{horiguchi2020end,horiguchi2022encoder}.

A parallel line of research has pursued non-autoregressive (NA) EEND models, which improve inference efficiency by extracting speaker attractors in parallel~\cite{rybicka2022end, fujita2023neural, hao2023end, chen2024attention, landini2024diaper}. To further refine intermediate representations, self-conditioning (SC)---originally proposed for CTC-based ASR~\cite{nozaki2021relaxing}---has been introduced into EEND, where intermediate diarization predictions $\hat{Y}_l$ are fed back as conditioning signals to subsequent encoder layers, enabling progressive refinement of speaker activity across the network depth~\cite{fujita2023neural, fujita2023self, landini2024diaper}.

However, a structural conflict exists between this SC mechanism and Permutation Invariant Training (PIT), the conventional training objective for resolving speaker-label permutation ambiguity. PIT's hard minimum greedily resolves permutations without constraining cross-iteration consistency, inducing \emph{label switching}---abrupt speaker index reassignments across training epochs---which has been shown to destabilize convergence in speech separation~\cite{huang2020stabilizing}. Applying PIT independently at each layer further introduces \emph{layer-decoupling}: adjacent layers freely adopt conflicting speaker assignments, as identified in deeper separation models~\cite{gao2021progressive} and addressed through cascaded training and layer-wise optimization in speech separation~\cite{yang2020interrupted, gao23e_interspeech}.

In EEND, auxiliary PIT losses at intermediate layers were first introduced by~\cite{yu2022auxiliary}, and further explored through layer-specific attention supervision~\cite{jeoung2023improving} and multi-resolution hierarchical fusion~\cite{kim25l_interspeech}; SC-based models inherit the same independent per-layer optimization~\cite{fujita2023neural, fujita2023self, landini2024diaper}, where each layer resolves its own permutation without cross-layer coordination. In SC-based EEND, however, this presents a more severe risk: because $\hat{Y}_l$ is directly passed as a conditioning signal to layer $l{+}1$, inter-layer permutation mismatches propagate as structured noise throughout the SC path---a failure mode that layer-wise independent optimization is inherently limited in addressing.

We formalize this as \emph{Hierarchical Permutation Inconsistency} (HPI): when each intermediate layer independently minimizes its own PIT loss, adjacent layers may resolve permutation ambiguity in a conflicting manner, propagating permutation noise into the SC conditioning path and disrupting progressive feature refinement.

To address HPI, we propose two complementary frameworks. \emph{Group-Wise PIT} (GW-PIT) imposes a hard constraint by tying all intermediate layers to a single global permutation, eliminating inter-layer mismatch. \emph{Continuous Influence PIT} (CI-PIT) applies a soft regularization scheme that aggregates cost matrices across layers via Gaussian weighting, promoting inter-layer permutation continuity while preserving layer-specific flexibility. To analyze how each framework affects permutation dynamics along the SC path, we further introduce three diagnostic metrics---permutation mismatch ratio (PMR), adjacent switching ratio (ASR), and epoch switching ratio (ESR)---and provide a systematic evaluation under each training objective.

\section{Methods}
\label{sec:methods}

This section defines the EEND-NA architecture with self-conditioning (SC)~\cite{fujita2023neural}, formulates the hierarchical permutation inconsistency problem arising from \textit{Layer-Wise PIT} (LW-PIT), and proposes \textit{Group-Wise PIT} (GW-PIT) and \textit{Continuous Influence PIT} (CI-PIT) to address this conflict.

\subsection{EEND-NA with Self-Conditioning}
\label{ssec:eend-na}

The EEND-NA model processes $X \in \mathbb{R}^{T \times D}$ through $L$ encoder 
blocks to estimate frame-wise speaker activity $\hat{Y} \in \mathbb{R}^{T \times S}$. 
Following~\cite{fujita2023neural}, a Multi-Head-Attention (MHA) module extracts 
intermediate attractors $A_l \in \mathbb{R}^{S \times D}$ in parallel, 
using $S$ learnable query vectors $Q \in \mathbb{R}^{S \times D}$:
\begin{equation}
    A_l = \text{MHA}(Q, E_l, E_l)
\end{equation}
The intermediate speaker probabilities $\hat{Y}_l = \text{Sigmoid}(E_l A_l^T)$ 
are re-injected as conditioning signals to subsequent layers:
\begin{equation}
    E_{l+1} = \text{EncoderBlock}_{l+1}
    \Bigl(E_l + \underbrace{\hat{Y}_l A_l W}_{\text{SC term}}\Bigr)
\end{equation}
where $W \in \mathbb{R}^{D \times D}$ is a single learnable weight matrix shared 
across all layers, and $\hat{Y}_l A_l W$ is the self-conditioning (SC) signal 
that injects layer-$l$ speaker predictions into the input of layer $l{+}1$.

\subsection{Problem Formulation: Layer-Wise PIT (LW-PIT)}

To resolve speaker-label permutation ambiguity, PIT selects the optimal permutation minimizing BCE:
\begin{equation}
    \mathcal{L}_{\text{PIT}}(Y, \hat{Y}) = \min_{\phi \in \text{Perm}(S)} \text{BCE}(Y^\phi, \hat{Y})
\end{equation}
where $Y^\phi \in \mathbb{R}^{T \times S}$ is the ground-truth permuted by $\phi$, and $\text{BCE}(\cdot, \cdot)$ denotes the binary cross-entropy aggregated over all $T$ frames and $S$ speakers. Yu et al.~\cite{yu2022auxiliary} introduced auxiliary PIT losses at intermediate layers, a scheme subsequently adopted by SC-based EEND studies~\cite{fujita2023neural, fujita2023self, landini2024diaper}. We re-formalize this as \textit{Layer-Wise PIT} (LW-PIT) with $\lambda_l = l/L$:
\begin{equation}
\label{eq:lw_pit}
    \mathcal{L}_{\text{LW-PIT}} = \sum_{l=1}^L \lambda_l \left( \min_{\phi_l \in \text{Perm}(S)} \text{BCE}(Y^{\phi_l}, \hat{Y}_l) \right)
\end{equation}
The structural problem lies in the position of the $\min$ operator: placed \textit{inside} the summation, it forces each layer to resolve its own optimal permutation $\phi_l$ independently. Under SC, this is particularly damaging: $\hat{Y}_l$ is passed to layer $l{+}1$ under permutation $\phi_l$, but layer $l{+}1$ independently resolves $\phi_{l+1}$. When $\phi_l \neq \phi_{l+1}$, the conditioning signal carries inconsistent speaker indices, propagating permutation noise into every subsequent layer and undermining the progressive refinement that SC is designed to achieve.

\subsection{Proposed Methods}

\subsubsection{Group-Wise PIT (GW-PIT)}

\textit{Group-Wise PIT} (GW-PIT) directly corrects the structural flaw of LW-PIT by enforcing a shared global permutation over $G \subseteq \{1, \dots, L\}$. The key modification is relocating the $\min$ operator \textit{outside} the summation:
\begin{equation}
\label{eq:gw_pit}
    \mathcal{L}_{\text{GW-PIT}} = \min_{\phi \in \text{Perm}(S)} \sum_{l \in G} \lambda_l \cdot \text{BCE}(Y^\phi, \hat{Y}_l)
\end{equation}
Rather than minimizing per-layer losses independently, $\mathcal{L}_{\text{GW-PIT}}$ searches for the single $\phi$ that minimizes the aggregate depth-weighted loss across all layers simultaneously. All layers in $G$ are therefore supervised under the same speaker-index assignment ($\phi = \phi_1 = \dots = \phi_{|G|}$), eliminating inter-layer mismatch by construction and ensuring that $\hat{Y}_l$ serves as a consistent conditioning signal throughout the SC path. In this work, we set $G = \{1, \dots, L\}$.

\begin{figure}[t]
  \centering
  \includegraphics[width=\linewidth]{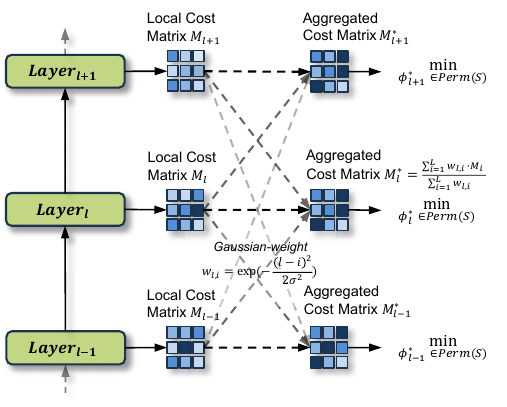}
  \caption{$\mathcal{L}_{\text{CI-PIT}}$ permutation assignment: Local cost matrices are aggregated via Gaussian weighting into a unified cost matrix, from which the optimal permutation $\phi_l^*$ is resolved via assignment minimization.}
  \label{fig:CI-PIT}
\end{figure}

\subsubsection{Continuous Influence PIT (CI-PIT)}

While $\mathcal{L}_{\text{GW-PIT}}$ eliminates HPI through a rigid global constraint, this hard coupling may overly restrict layer-specific gradients, limiting representational flexibility particularly in early training. \textit{Continuous Influence PIT} (CI-PIT) provides a softer alternative: rather than imposing a single permutation, it shapes each layer's permutation resolution by aggregating cost information from neighboring layers using Gaussian weighting. This encourages inter-layer permutation continuity without discarding layer-specific signals. Extending~\cite{lin2020optimal}, CI-PIT proceeds in three steps, as illustrated in Figure~\ref{fig:CI-PIT}.

\textbf{Step 1: Gaussian Kernel Weighting.}
Let $l$ denote the target layer and $i$ index all layers 
contributing to its aggregated cost matrix. The influence weight of layer $i$ on layer $l$ is:
\begin{equation}
    w_{l,i} = \exp\left(-\frac{(l-i)^2}{2\sigma^2}\right)
\end{equation}
where $\sigma$ controls the influence range. Layers close in depth contribute strongly; distant layers contribute exponentially less.

\textbf{Step 2: Weighted Aggregation of Cost Matrices.}
Let $M_i \in \mathbb{R}^{S \times S}$ be the pairwise BCE cost matrix at layer $i$. The aggregated matrix for layer $l$ is:
\begin{equation}
    M_l^* = \frac{\sum_{i=1}^L w_{l,i} M_i}{\sum_{i=1}^L w_{l,i}}
\end{equation}

\textbf{Step 3: Permutation Resolution and Loss Computation.}
Following~\cite{lin2020optimal}, which resolves speaker assignment via the Hungarian algorithm~\cite{jonker1986improving} on a pairwise BCE cost matrix in $\mathcal{O}(N^3)$ time, we apply the same procedure to $M_l^*$:
\begin{equation}
    \phi_l^* = \text{Hungarian}(M_l^*)
\end{equation}
\begin{equation}
\label{eq:ci_pit}
    \mathcal{L}_{\text{CI-PIT}} = \sum_{l=1}^L \lambda_l \cdot \text{BCE}(Y^{\phi_l^*}, \hat{Y}_l)
\end{equation}
By varying $\sigma$, $\mathcal{L}_{\text{CI-PIT}}$ interpolates 
between $\mathcal{L}_{\text{LW-PIT}}$ ($\sigma \to 0$) and 
$\mathcal{L}_{\text{GW-PIT}}$ ($\sigma \to \infty$), making 
CI-PIT a unified framework that subsumes both as special cases.

\subsection{Permutation Consistency Metrics}

To quantify permutation variation along the layer and epoch axes, we define three metrics using $\mathbb{I}(\cdot)$ over $N$ training samples:
\begin{itemize}
    \item \textbf{Permutation Mismatch Ratio (PMR):} The ratio of mismatch between the optimal permutation of layer $l$ and the final layer $L$.
    \begin{equation}
        \text{PMR}(l) = \frac{1}{N} \sum_{n=1}^N \mathbb{I}(\phi_l^{(n)} \neq \phi_L^{(n)})
    \end{equation}
    \item \textbf{Adjacent Switching Ratio (ASR):} The rate of permutation change between adjacent layers ($l, l+1$), directly representing the noise level injected into the SC path.
    \begin{equation}
        \text{ASR}(l) = \frac{1}{N} \sum_{n=1}^N \mathbb{I}(\phi_l^{(n)} \neq \phi_{l+1}^{(n)})
    \end{equation}
    \item \textbf{Epoch Switching Ratio (ESR):} The frequency of permutation changes within the same layer $l$ across consecutive training epochs $e$ and $e+1$.
    \begin{equation}
        \text{ESR}(l) = \frac{1}{N} \sum_{n=1}^N \mathbb{I}(\phi_{l,e}^{(n)} \neq \phi_{l,e+1}^{(n)})
    \end{equation}
\end{itemize}

\section{Experimental Setup}
\label{sec:exp}

\subsection{Datasets}
\label{ssec:dataset}

\begin{table}[th]
  \caption{Statistics of datasets. For CALLHOME, the Length and Overlap values are reported for the Adaptation / Test splits, respectively.}
  \label{tab:dataset}
  \centering
  \footnotesize 
  \resizebox{\linewidth}{!}{
  \begin{tabular}{@{} l c c c c @{}}
    \toprule
    \textbf{Dataset} & \textbf{Num. Spk.} & \textbf{Split} & \textbf{Length (h)} & \textbf{Overlap (\%)} \\
    \midrule
    \multirow{2}{*}{\textbf{SimConv}} & 2 & Train & 2,478 & 7.2 \\
                                      & 3 & Train & 2,513 & 11.3 \\
    \midrule
    \multirow{2}{*}{\textbf{CALLHOME}} & 2 & Adapt / Test & 3.19 / 2.97 & 14.0 / 13.1 \\
                                       & 3 & Adapt / Test & 2.11 / 2.98 & 19.6 / 17.0 \\
    \bottomrule
  \end{tabular}
  \vspace{-5mm}
    }
\end{table}

We evaluated our proposed frameworks on the CALLHOME(NIST SRE 2000) dataset \cite{doddington2000nist}, a standard benchmark for speaker diarization. We conducted experiments on both 2-speaker and 3-speaker scenarios. The dataset was partitioned into adaptation and test sets using the standard CALLHOME partition \footnote{\url{https://github.com/BUTSpeechFIT/CALLHOME_sublists}}. 
For training, we generated simulated conversations (SimConv)\footnote{\url{https://github.com/BUTSpeechFIT/EEND_dataprep/tree/main/v2/LibriSpeech}} \cite{landini2022simulated,landini2023multi} using the LibriSpeech corpus \cite{panayotov2015librispeech} as the source speech. To match the acoustic conditions of the CALLHOME dataset, the original 16~kHz audio was downsampled to 8~kHz. To simulate realistic overlap patterns and utterance distributions, we extracted statistical priors from the CALLHOME adaptation set for the 2-speaker and 3-speaker conditions, respectively. Following previous work~\cite{landini2023multi}, we augmented the training data with MUSAN noise \cite{snyder2015musan} without applying room impulse responses (RIR). Detailed dataset statistics are summarized in Table~\ref{tab:dataset}.

\subsection{Model Configurations}
\label{ssec:config}
Our baseline architecture follows the EEND-NA~\cite{fujita2023neural} described
in Section~\ref{ssec:eend-na}, re-implemented using the PyTorch-based
EEND-EDA codebase\footnote{\url{https://github.com/BUTSpeechFIT/EEND}}
and trained from scratch on our SimConv datasets.
The input features are 23-dimensional log Mel-filterbanks (10~ms shift),
stacked over 15 consecutive frames into a 345-dimensional input vector
every 100~ms. The Transformer encoder consists of 8 layers with a
256-dimensional embedding size, 4 attention heads, and a 2048-dimensional
FFN. We used 500-frame audio segments with batch size 64, optimized
with Adam~\cite{kingma2014adam} and a Noam learning rate
scheduler~\cite{vaswani2017attention} (100k warm-up steps).
Pre-training on SimConv was performed for 100 epochs; the model was
subsequently fine-tuned on the CALLHOME adaptation set for another
100 epochs at a fixed learning rate of $10^{-5}$. In both stages,
the weights of the last 10 epoch checkpoints were averaged to obtain
the final model.
During inference, the full recording is processed in a single forward
pass. DER is evaluated with a 0.25~s forgiveness collar; a median
filter with a window size of 11 is applied as post-processing over
the network's output, and speaker activities are determined by
thresholding the output probabilities at 0.5.

\section{Results and Analysis}
\label{sec:results}

\subsection{Impact on Diarization Performance}
\label{ssec:main_der}
\begin{table}[th]
  \caption{Performance comparison of different PIT objectives on the EEND-NA model with Self-Conditioning. All values in \%.}
  \label{tab:main_der}
  \centering
  \resizebox{\linewidth}{!}{
  \begin{tabular}{lcccccccc}
    \toprule
    \multirow{2}{*}{\textbf{Training Objective}} & \multicolumn{4}{c}{\textbf{CALLHOME 2-spk}} & \multicolumn{4}{c}{\textbf{CALLHOME 3-spk}} \\
    \cmidrule(lr){2-5} \cmidrule(lr){6-9}
     & \textbf{MI} & \textbf{FA} & \textbf{CF} & \textbf{DER} & \textbf{MI} & \textbf{FA} & \textbf{CF} & \textbf{DER} \\
    \midrule
    $\mathcal{L}_{\text{LW-PIT}}$ (Baseline) & \textbf{4.11} & 3.65 & \textbf{0.53} & 8.30 & 7.73 & 2.23 & 3.13 & 13.04 \\
    $\mathcal{L}_{\text{GW-PIT}}$ & 4.21 & 3.17 & 0.64 & 8.02 & 8.42 & 2.18 & 3.06 & 13.66 \\
    $\mathcal{L}_{\text{CI-PIT}}$($\sigma$=1.0) & 4.36 & \textbf{2.59} & 0.57 & \textbf{7.52} & \textbf{7.62} & \textbf{2.07} & \textbf{2.83} & \textbf{12.52} \\
    \bottomrule
  \end{tabular}
  \vspace{-4mm}
  }
\end{table}
Table~\ref{tab:main_der} presents DER and its decomposition---Missed Speech (MI), False Alarm (FA), and Speaker Confusion (CF)---under each PIT objective on CALLHOME 2- and 3-speaker sets.
In the 2-speaker scenario, $\mathcal{L}_{\text{GW-PIT}}$ provides a marginal improvement over $\mathcal{L}_{\text{LW-PIT}}$ (8.02\% vs.\ 8.30\%), while $\mathcal{L}_{\text{CI-PIT}}$ achieves a substantially lower DER of 7.52\%, driven primarily by a large reduction in FA (3.65\%~$\to$~2.59\%). In the more challenging 3-speaker scenario, $\mathcal{L}_{\text{GW-PIT}}$ degrades performance relative to the baseline, whereas $\mathcal{L}_{\text{CI-PIT}}$ delivers the best DER of 12.52\%---the only method with consistent gains across both settings. These results suggest that the rigid global constraint of $\mathcal{L}_{\text{GW-PIT}}$ becomes detrimental as number of speakers increases, while the soft, localized alignment of $\mathcal{L}_{\text{CI-PIT}}$ scales more effectively.

\subsection{Training Dynamics of Permutation Consistency}
\label{ssec:dynamics}

\begin{figure}[t]
  \centering
  \includegraphics[width=\linewidth]{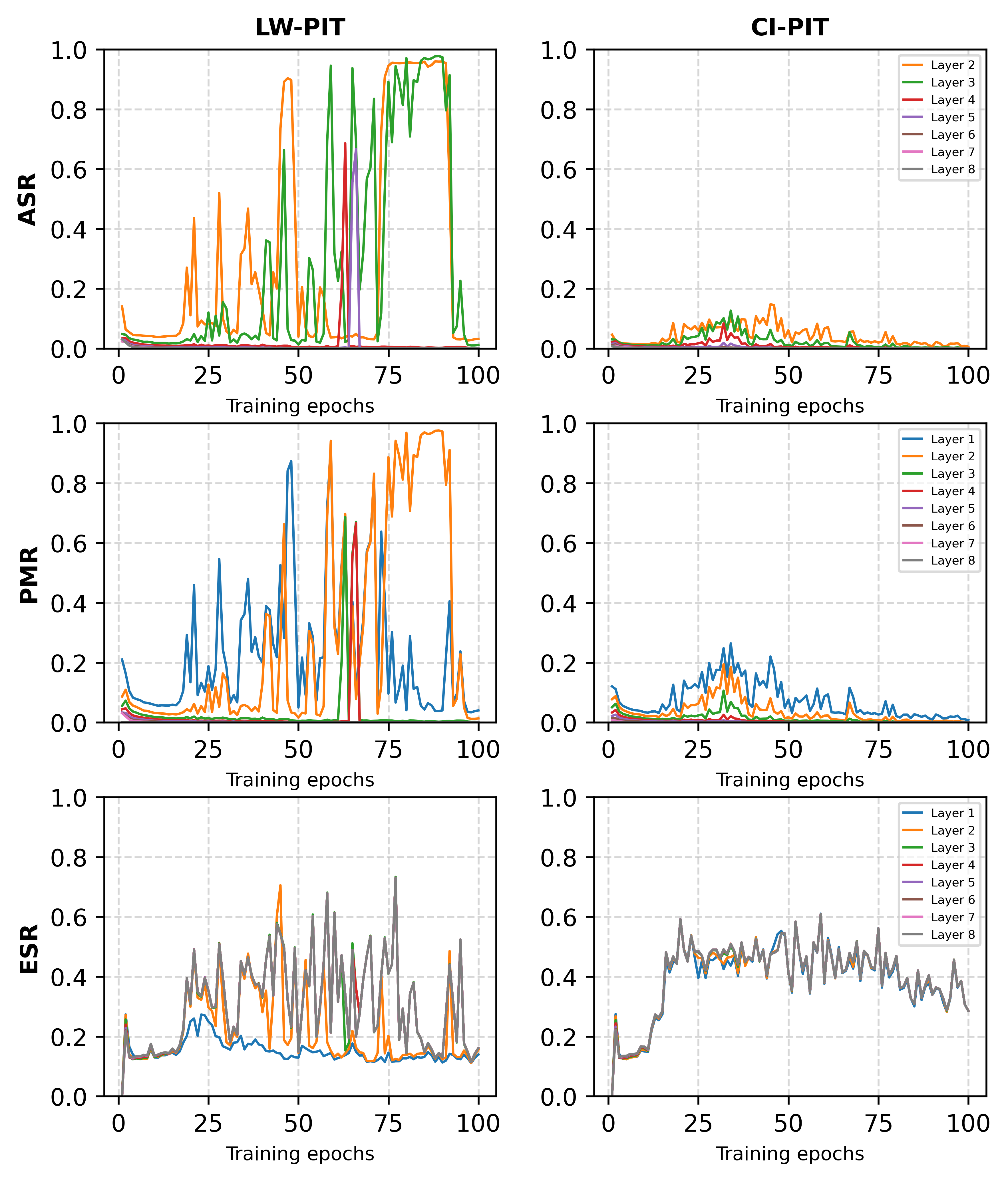}
  \caption{Training dynamics of permutation consistency metrics on CALLHOME 2-spk. ASR, PMR, and ESR are plotted over epochs for each encoder layer under $\mathcal{L}_{\text{LW-PIT}}$ (left column) and $\mathcal{L}_{\text{CI-PIT}}$ (right column).}
  \label{fig:training_dynamics}
  \vspace{-2mm}
\end{figure}

Figure~\ref{fig:training_dynamics} tracks the epoch-wise progression 
of ASR, PMR, and ESR per encoder layer under $\mathcal{L}_{\text{LW-PIT}}$ 
and $\mathcal{L}_{\text{CI-PIT}}$, offering direct empirical evidence for HPI.

Under $\mathcal{L}_{\text{LW-PIT}}$, all three metrics remain persistently 
high and volatile. Elevated ASR confirms that adjacent layers frequently 
adopt conflicting permutations, corrupting the SC conditioning path. 
Large PMR in lower layers reveals systematic misalignment with the final 
layer, while asynchronous ESR spiking across epochs reflects independently 
unstable assignments uncoordinated across the network depth.

Under $\mathcal{L}_{\text{CI-PIT}}$, ASR and PMR converge near zero after 
an initial exploration phase, confirming inter-layer alignment and consistent 
speaker labeling with the final layer. ESR stabilizes at a moderate level 
with tightly coupled trajectories across layers---permutation changes persist, 
but occur in coordination rather than asynchronously, in contrast to the 
instability observed under $\mathcal{L}_{\text{LW-PIT}}$.

\subsection{Hierarchical Permutation Consistency at Convergence}
\label{ssec:esr_l1}

\begin{table}[t]
  \caption{ESR $L_1$ distance between each intermediate layer ($l$) and the final layer ($L=8$) on CALLHOME 2-spk, defined as the sum of absolute differences between their epoch-wise ESR values across all training epochs (i.e., $\sum_e |\text{ESR}_e(l) - \text{ESR}_e(L)|$).}
  \label{tab:esr_l1}
  \centering
  \footnotesize
  \begin{tabular}{lccc}
    \toprule
    \textbf{Layer pair} & $\mathcal{L}_{\text{LW-PIT}}$ & $\mathcal{L}_{\text{GW-PIT}}$ & $\mathcal{L}_{\text{CI-PIT}}$ \\
    \midrule
    1 vs 8 & \textcolor{red}{15.68} & 0.00 & \textcolor{blue}{1.24} \\
    2 vs 8 & \textcolor{red}{9.27}  & 0.00 & \textcolor{blue}{0.56} \\
    3 vs 8 & 0.63  & 0.00 & 0.19 \\
    4 vs 8 & 0.27  & 0.00 & 0.09 \\
    5 vs 8 & 0.05  & 0.00 & 0.05 \\
    6 vs 8 & 0.03  & 0.00 & 0.02 \\
    7 vs 8 & 0.02  & 0.00 & 0.01 \\
    \bottomrule
  \end{tabular}
  \vspace{-4mm}
\end{table}

Table~\ref{tab:esr_l1} reports the $L_1$ distance of ESR between each layer and the final layer (Layer~8) at convergence. Under $\mathcal{L}_{\text{LW-PIT}}$, lower layers exhibit extreme distances (15.68 and 9.27 for pairs 1~vs~8 and 2~vs~8), confirming severe HPI. $\mathcal{L}_{\text{GW-PIT}}$ eliminates all distances by construction, yet this perfect fixation yields only marginal gains in the 2-speaker case and degrades 3-speaker performance (Table~\ref{tab:main_der}), indicating that zero flexibility is sub-optimal. $\mathcal{L}_{\text{CI-PIT}}$ strikes the optimal balance: substantially reducing lower-layer distances (1.24 and 0.56) while preserving small but non-zero values in deeper layers, retaining the representational flexibility necessary for hierarchical feature extraction.

\subsection{Interaction with Self-Conditioning}
\label{ssec:ablation_sc}

\begin{table}[t]
  \caption{Ablation study on SC and PIT variants on CALLHOME 2-spk. All values in \%.}
  \label{tab:ablation_sc_pit}
  \centering
  \footnotesize
  \begin{tabular}{lcccc}
    \toprule
    \textbf{Method} & \textbf{MI} & \textbf{FA} & \textbf{CF} & \textbf{DER} \\
    \midrule
    $\mathcal{L}_{\text{LW-PIT}}$ & 4.94 & \textbf{2.54} & 0.83 & 8.34 \\
    \hspace{3mm} + SC & \textbf{4.11} & 3.65 & \textbf{0.53} & 8.30 \\
    \midrule
    $\mathcal{L}_{\text{GW-PIT}}$ & 4.83 & 2.81 & 0.86 & 8.51 \\
    \hspace{3mm} + SC & 4.21 & 3.17 & 0.64 & 8.02 \\
    \midrule
    $\mathcal{L}_{\text{CI-PIT}}$ & 4.88 & 2.80 & 0.99 & 8.67 \\
    \hspace{3mm} + SC & 4.36 & 2.59 & 0.57 & \textbf{7.52} \\
    \bottomrule
  \end{tabular}
  \vspace{-3mm}
\end{table}

Table~\ref{tab:ablation_sc_pit} isolates the contribution of SC under each PIT objective. Adding SC to $\mathcal{L}_{\text{LW-PIT}}$ yields negligible improvement (8.34\%~$\to$~8.30\%), as permutation volatility corrupts the conditioning path. $\mathcal{L}_{\text{GW-PIT}}$ benefits more from SC (8.51\%~$\to$~8.02\%), yet remains limited by its over-constrained intermediate representations. The combination of $\mathcal{L}_{\text{CI-PIT}}$ and SC yields the largest reduction (8.67\%~$\to$~7.52\%), despite $\mathcal{L}_{\text{CI-PIT}}$ alone underperforming the baseline. This synergy demonstrates that $\mathcal{L}_{\text{CI-PIT}}$ is specifically designed to unlock SC's potential: by suppressing permutation noise without eliminating layer-specific flexibility, it enables the progressive refinement that SC is intended to achieve.

\subsection{Effect of Gaussian Scope in CI-PIT}
\label{ssec:sigma}

\begin{table}[t]
  \caption{Effect of the Gaussian width $\sigma$ in $\mathcal{L}_{\text{CI-PIT}}$ on CALLHOME 2-spk. All values in \%.}
  \label{tab:sigma}
  \centering
  \footnotesize
  \begin{tabular}{ccccc}
    \toprule
    $\sigma$ & \textbf{MI} & \textbf{FA} & \textbf{CF} & \textbf{DER} \\
    \midrule
    1.0 & 4.36 & \textbf{2.59} & 0.57 & \textbf{7.52} \\
    2.0 & \textbf{4.09} & 3.47 & \textbf{0.42} & 7.97 \\
    3.0 & 4.18 & 3.29 & 0.66 & 8.12 \\
    \bottomrule
  \end{tabular}
  \vspace{-6mm}
\end{table}

Table~\ref{tab:sigma} shows that $\sigma = 1.0$ yields the best DER of 7.52\%. As $\sigma$ increases, the aggregation scope widens and $\mathcal{L}_{\text{CI-PIT}}$ approaches the behavior of $\mathcal{L}_{\text{GW-PIT}}$, with DER degrading to 7.97\% and 8.12\% at $\sigma = 2.0$ and $3.0$, respectively. This confirms that localized, layer-proximate permutation alignment is optimal: $\sigma$ controls the trade-off between consistency and flexibility, and excessive coupling eliminates the layer-specific differentiation essential for modeling complex speaker transitions.

\section{Conclusions}
\label{sec:conclusion}

This paper identified that applying independent per-layer PIT---formalized as \textit{Layer-Wise PIT} ($\mathcal{L}_{\text{LW-PIT}}$)---to self-conditioned EEND models causes \textit{Hierarchical Permutation Inconsistency} (HPI), injecting structured noise into the SC path and disrupting progressive refinement. To address this, we proposed \textit{Group-Wise PIT} ($\mathcal{L}_{\text{GW-PIT}}$) and \textit{Continuous Influence PIT} ($\mathcal{L}_{\text{CI-PIT}}$), which softly aligns permutations across the layer hierarchy via Gaussian-weighted cost matrix aggregation. Evaluations on CALLHOME show that $\mathcal{L}_{\text{CI-PIT}}$ suppresses permutation noise while preserving representational flexibility, achieving consistent DER improvements across 2- and 3-speaker scenarios.

\ifcameraready
\section{Acknowledgments}
This work was supported by KOITA grant funded by MSIT(KOITA20250002-38).
\fi

\section{Generative AI Use Disclosure}
Google Gemini was used to assist with English translation, language refinement, and partial code development during the preparation of this manuscript. All research ideas, problem formulations, experimental designs, and result analyses were conducted solely by the authors. All AI-assisted content was reviewed and verified under the authors' supervision.

\bibliographystyle{IEEEtran}
\bibliography{mybib}

\end{document}